\documentclass[10pt, conference, letterpaper]{IEEEtran}
\IEEEoverridecommandlockouts
\usepackage{ifpdf}
\usepackage{color}
\usepackage[latin1]{inputenc}
\usepackage{cite}
\usepackage{mathtools}
\usepackage{amssymb}
\usepackage{url,cite}

\usepackage{tikz}
\usetikzlibrary{decorations.pathmorphing} % noisy shapes
\usetikzlibrary{fit}					% fitting shapes to coordinates
\usetikzlibrary{backgrounds}	% drawing the background after the foreground
\usetikzlibrary{shapes}
\usetikzlibrary{positioning}
\usetikzlibrary{shadows}
\usetikzlibrary{decorations.pathmorphing}
\usetikzlibrary{decorations.shapes}
\usepackage{multirow}
\usepackage{dcolumn}
\usepackage{pgfplots}
\pgfplotsset{compat=1.3}% <-- moves axis labels near ticklabels (respects tick label widths)

\newtheorem{lemma}{Lemma}

\begin{document}
\allowdisplaybreaks

\title{\huge Outage Duration in Poisson Networks and its Application to Erasure Codes
\thanks{This work was supported by the Austrian Science Fund (FWF) under grant P24480-N15 and by the K-project DeSSnet (Dependable, secure and time-aware sensor networks), which is funded within the context of COMET -- Competence Centers for Excellent Technologies by the Austrian Ministry for Transport, Innovation and Technology (BMVIT), the Federal Ministry for Digital and Economic Affairs (BMDW), and the federal states of Styria and Carinthia; the COMET program is conducted by the Austrian Research Promotion Agency~(FFG).}}
\author{Udo Schilcher$^{1,2}$, Siddhartha Borkotoky$^1$, Jorge F. Schmidt$^2$, and Christian Bettstetter$^{1,2}$,\\
$^1$ Lakeside Labs GmbH, Klagenfurt, Austria\\
$^2$ University of Klagenfurt, Austria\\
email: \texttt{udo.schilcher@aau.at}
}

% notation
\newcommand{\prob}[1]{{\mathbb{P}\!\left[{#1}\right]}}
\newcommand{\expect}[1]{{{\mathbb{E}}\!\left[{#1}\right]}}
\newcommand{\expecto}{{\mathbb{E}}}
\newcommand{\var}[1]{{{\rm var}\!\left[{#1}\right]}}
\newcommand{\cov}[2]{{{\rm cov}\!\left[{#1},{#2}\right]}}
\newcommand{\cor}[2]{{\rho\!\left[{#1},{#2}\right]}}

\newcommand{\R}{\mathbb{R}}
\newcommand{\N}{\mathbb{N}}
\newcommand{\dd}{{\rm d}}
\newcommand{\uniform}{\mathbb{U}(0,1)}

\newcommand{\ppp}{\Phi}
\newcommand{\interf}{I}
\newcommand{\fad}[1]{h^2_{#1}}
\newcommand{\ploss}[1]{\ell\left(\left\|#1\right\|\right)}
\newcommand{\plosssc}[1]{\ell\left(#1\right)}
\newcommand{\plosssq}[1]{\ell^2(\|#1\|)}
\newcommand{\channel}[2]{\fad{#1}{#2}\,\ploss{#1}}
\newcommand{\tx}[1]{\gamma_{#1}}
\newcommand{\txone}[1]{\gamma^*_{#1}}
\newcommand{\txtwo}[1]{\gamma^{**}_{#1}}
\newcommand{\prdensii}[1]{\rho^{(2)}(#1)}
\newcommand{\dens}{\lambda}

\newcommand{\txp}{p}
\newcommand{\rxp}{p_x}
\newcommand{\nakm}{m}
\newcommand{\plc}{\alpha}

\newcommand{\suc}[1]{{\prob{S\geq#1}}}
\newcommand{\sucex}[1]{{\prob{S=#1}}}
\newcommand{\out}[1]{{\prob{O\geq#1}}}
\newcommand{\outex}[1]{{\prob{O=#1}}}
\newcommand{\suckn}[2]{{\prob{\sucn{#1}=#2}}}
\newcommand{\sucn}[1]{S(#1)}

\newcommand{\divpoly}[1]{D_{#1}(p,\delta)}

\newcommand{\sir}{\mathrm{SIR}}

\newcommand{\through}{\Omega}

\newcommand{\decp}{\mathbb{P}_\mathrm{dec}(m,k)}

\sloppy
\markboth{Schilcher et al.: Outage Duration in Poisson Networks and its Application to Erasure Codes}{aa}

\maketitle
\thispagestyle{empty}
\pagestyle{headings}

\begin{abstract}
We derive the probability distribution of the link outage duration at a typical receiver in a wireless network with Poisson distributed interferers sending messages with slotted random access over a Rayleigh fading channel. This result is used to analyze the performance of random linear network coding, showing that there is an optimum code rate and that interference correlation affects the decoding probability and throughput.
\end{abstract}

\section{Introduction}

The analytical modeling of link outage in wireless networks should consider the fact that interference is correlated. Such correlation leads to the effect that time slots being in outage tend to cluster together, i.e., they come in bursts~\cite{atiq19:eusipco}. This letter analyzes this effect using stochastic geometry. For Poisson networks with small-scale fading and slotted random access, we derive stochastic expressions for the outage duration and the duration between two outages, called success period. These expressions allow us to quantify how outage and success periods become longer with increasing~correlation.

Results are relevant for the design of transmission schemes, including diversity~\cite{yang2003average2}, equalization~\cite{rugini2005simple}, and channel coding with interleaving~\cite{shi2004interleaving}.
To show the effects by an example, we analyze the performance of erasure correction coding under the impact of correlated interference. 
Correlated interference and the resulting clustering of outage events degrades the performance of coding for the following reason: Let us assume that a code can compensate for a certain number of lost out of a given number of coded packets. Without interference correlation we would on average lose a few packets but the code can compensate these losses.
When having clustering of outage, however, sometimes all coded packets are received while at other times too few are received, and the code cannot always compensate these losses.

In particular, we compute the decoding probability and throughput of random linear network coding~\cite{1705002}. We observe that there is an optimal amount of redundancy due to the tradeoff between weak code performance (in case of low redundancy) and high interference (in case of high redundancy). Furthermore, we find that the correlation of interference impacts the throughput of a typical node in the network and it depends on the specific scenario whether throughput is increased or decreased.

Results are relevant for devices with constraints in terms of computational power or energy such as industrial wireless sensor networks~\cite{borkotoky19:globecom}. In such a setup, random linear network coding is an interesting option as, depending on the field size adopted, it can be very efficient while still being effective. The provided tools can be applied to analyze the performance of the network and optimize the amount of redundancy depending on the particular network setup. Practical studies in an industrial sensor network within facilities of a project partner from process industry are planned to further pursue this direction of research.

\section{Network model}

Nodes are distributed in space according to a Poisson point process (PPP) $\Phi\subseteq\R^2$ with intensity $\lambda$. Time is slotted, and each node transmits in each slot i.i.d. with probability $p$.
A typical receiver can be located at the origin $o$ due to Slyvnjak's theorem~\cite{haenggi13:book}. It aims to receive messages from a sender located at $s$ that transmits in each slot, i.e., with sending probability 1, and is not part of $\Phi$.

The wireless channel is modeled by a distance dependent path loss combined with fading. The interference power arriving at $o$ from a node $x\in\Phi$ is $\rxp=\kappa\,\|x\|^{-\alpha}\,\fad{x}\,\tx{x}$,
where $\kappa$ is the transmission power, $\alpha$ is the path loss exponent, $\fad{x}$ is the channel gain modeling multi-path propagation following an exponential distribution with unit mean for Rayleigh fading, and $\tx{x}$ is the indicator function determining whether or not $x$ transmits in the current slot. We assume that fading is independent over time slots and space.

The signal-to-interference ratio (SIR) at $o$ is
\begin{equation}
\sir=\frac{\|s\|^{-\alpha}\fad{s}}{\sum_{x\in\Phi}\|x\|^{-\alpha}\fad{x}\tx{x}}\:.
\end{equation}
A transmission is assumed to be received correctly if $\sir>\theta$ for a given threshold $\theta$. The specific value of $\theta$ depends on properties of the receiver.

\section{Success and Outage Durations}\label{sec:exp}
\subsection{Duration between outages}
The duration between consecutive outages, or equivalently the duration of success $S$, at a typical receiver increases with increasing interference correlation.
\begin{lemma}
The probability mass function (pmf) of the duration between outages $S$ is 
\begin{equation}
\sucex{n}=\exp\big(-\Delta\divpoly{n})-\exp(-\Delta\divpoly{n+1}\big)\:,
\end{equation}
where $\Delta=\lambda\pi s^2\theta^\delta \Gamma(1+\delta)\Gamma(1-\delta)$, $\delta=\frac{2}{\alpha}$, and $\divpoly{n}=\sum_{k=1}^n{n\choose k}{\delta-1\choose k-1}p^k$ is the $n$th diversity polynomial~\cite{haenggi13:div-poly}.
\end{lemma}
\begin{IEEEproof}
The probability that a receiver can correctly receive all messages in $n$ consecutive slots is
\begin{eqnarray}\label{eq:succ}
\suc{n}&=&\prob{\sir_1\geq\theta,\dots,\sir_n\geq\theta}\\\nonumber
&\stackrel{(a)}{=}&\expect{\exp\left(-\frac{\theta\sum_{x\in\Phi}\|x\|^{-\alpha}\fad{x}\tx{x}}{\|s\|^{-\alpha}}\right)^n}\\\nonumber
&=&\exp\big(-\Delta\divpoly{n}\big)\:,
\end{eqnarray}
where $(a)$ holds due to the independence of the random fading gains $h_x^2$ in different slots.
The probability that the success duration is $S=n$ slots is then calculated by $\sucex{n}=\suc{n}-\suc{n+1}$ and substituting~\eqref{eq:succ} twice.
\end{IEEEproof}

It follows that the expected success duration is
\begin{equation}
\expect{S}=\sum_{n=1}^\infty n\,\sucex{n}=\sum_{n=1}^\infty \suc{n}\:
\end{equation}
and its variance is
\begin{equation}
\var{S}=\sum_{n=1}^\infty n^2\,\sucex{n}=\sum_{n=1}^\infty (2n-1)\,\suc{n}\:.
\end{equation}

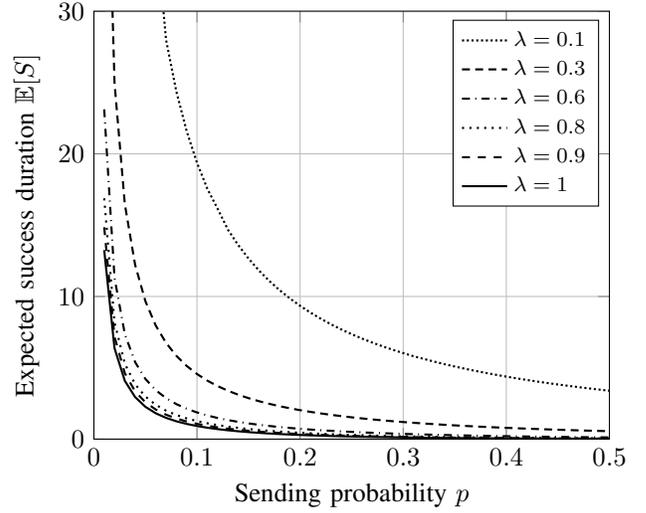
\begin{figure}[t]
\centering
\begin{tikzpicture}
\begin{axis}[xlabel={Sending probability $p$},ylabel={Expected success duration $\expect{S}$},ymin=0,ymax=30,xmin=0,xmax=0.5,grid=both,
legend style={at={(0.98,0.98)}, anchor=north east, font=\footnotesize},
legend cell align=left
]

\addplot plot[color=black,densely dotted,no marks,style=thick]  table[x index=0,y index=1]  {succdur_lam_p.txt};\addlegendentry{$\lambda=0.1$}
\addplot plot[color=black,densely dashed,no marks,style=thick]  table[x index=0,y index=3]  {succdur_lam_p.txt};\addlegendentry{$\lambda=0.3$}
\addplot plot[color=black,dashdotted,no marks,style=thick]  table[x index=0,y index=6]  {succdur_lam_p.txt};\addlegendentry{$\lambda=0.6$}
\addplot plot[color=black,dotted,no marks,style=thick]  table[x index=0,y index=8]  {succdur_lam_p.txt};\addlegendentry{$\lambda=0.8$}
\addplot plot[color=black,dashed,no marks,style=thick]  table[x index=0,y index=9]  {succdur_lam_p.txt};\addlegendentry{$\lambda=0.9$}
\addplot plot[color=black,solid,no marks,style=thick]  table[x index=0,y index=10]  {succdur_lam_p.txt};\addlegendentry{$\lambda=1$}

\end{axis}
\end{tikzpicture}
\caption{Expected success duration over sending probability $p$, for different $\lambda$. Parameters are $\theta=1$, $\alpha=3$, and $s=(1,0)$. Note that the glitches are due to numerical instability of the calculations.}\label{fig:succdur_lam_p}
\end{figure}

We now study the effects of the system parameters on the success duration. For this purpose, we fix $s=(1,0)$, $\kappa=1$, and $\theta=1$, and vary $\lambda$, $\alpha$, and $p$.
Fig.~\ref{fig:succdur_lam_p} plots $\expect{S}$ over the sending probability $p$ for different interferer intensities $\lambda$.
We see that the expected success duration decreases rapidly when increasing $p$ or $\lambda$. In the limits, it approaches infinity for $\lambda\to 0$ (or for $p\to 0$) and zero for $\lambda\to\infty$ and any positive~$p$.

\begin{figure}[t]
\centering
\begin{tikzpicture}
\begin{axis}[xlabel={Interference correlation $\rho$},ylabel={Expected success duration $\expect{S}$},ymin=0,ymax=40,xmin=0,xmax=0.25,grid=both,
legend style={at={(0.98,0.98)}, anchor=north east, font=\footnotesize},
legend cell align=left,
xticklabel style={
        /pgf/number format/fixed,
        /pgf/number format/precision=2
}
]

\addplot plot[color=black,solid,no marks,style=thick]  table[x index=0,y index=10]  {succdur_alpha.txt};
\addplot plot[color=black,solid,no marks,style=thick]  table[x index=0,y index=9]  {succdur_alpha.txt};
\addplot plot[color=black,solid,no marks,style=thick]  table[x index=0,y index=8]  {succdur_alpha.txt};
\addplot plot[color=black,solid,no marks,style=thick]  table[x index=0,y index=5]  {succdur_alpha.txt};
\addplot plot[color=black,solid,no marks,style=thick]  table[x index=0,y index=3]  {succdur_alpha.txt};
\addplot plot[color=black,solid,no marks,style=thick]  table[x index=0,y index=1]  {succdur_alpha.txt};

\node at(axis cs:0.103,29) {$\alpha=3$};
\node at(axis cs:0.13,25) {$2.9$};
\node at(axis cs:0.14,21) {$2.8$};
\node at(axis cs:0.153,12) {$2.5$};
\node at(axis cs:0.157,7) {$2.3$};
\node at(axis cs:0.159,3) {$2.1$};

\end{axis}
\end{tikzpicture}
\caption{Expected success duration over interference correlation $\rho=\frac{p}{2}$, keeping the success probability $\suc{1}$ constant by having $\lambda p=0.01$. Parameters are $\theta=1$ and $s=(1,0)$.}\label{fig:succdur_p_lam}
\end{figure}
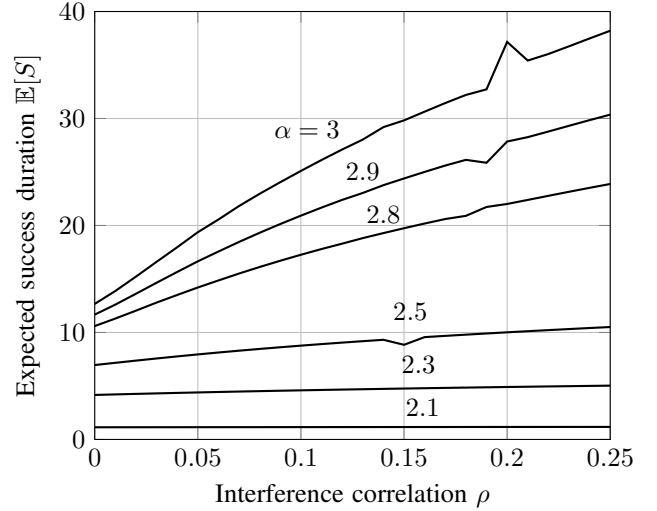

Fig.~\ref{fig:succdur_p_lam} studies the impact of interference correlation $\rho=\frac{p}{2}$~\cite{ganti09:interf-correl} on the expected success duration. The result is not diverted by the effect that higher interference leads to shorter success periods; this is achieved by varying $p$ while keeping the product $\lambda p$ and in turn the success probability $\suc{1}$ constant.
As can be seen, the success duration increases monotonically with $\rho$.
This overall increase is stronger for higher $\alpha$, with the impact of nearby interferers being stronger than those of distant ones. 
In the limit $p\to 0$ with constant $p\lambda$, we reach the case of uncorrelated interference~($\rho\to 0$).

\subsection{Duration of outages}
We study the outage duration as a function of the success duration probabilities $\suc{n}$.
\begin{lemma}\label{lem:pmf-out}
The pmf of the outage duration $O$ is 
\begin{equation}
\outex{n}=\sum_{k=0}^n {n\choose k} (-1)^k \:\suc{k+1}\:.
\end{equation}
\end{lemma}
\begin{IEEEproof}
The probability of $n$ consecutive outage slots is
\begin{eqnarray}
\out{n}\!\!&=&\!\!\expect{\left(1-\exp\left(-\frac{\theta\sum_{x\in\Phi}\|x\|^{-\alpha}\fad{x}\tx{x}}{\|s\|^{-\alpha}}\right)\right)^n}\nonumber\\
&=&\!\!\sum_{k=0}^n {n\choose{k}} (-1)^k \suc{k}\:.
\end{eqnarray}
Hence, the probability of an outage duration $n$ is
\begin{eqnarray}
\outex{n}&=&\mathbb{E}\Bigg[\left(1-\exp\left(-\frac{\theta\sum_{x\in\Phi}\|x\|^{-\alpha}\fad{x}\tx{x}}{\|s\|^{-\alpha}}\right)\right)^n\nonumber\\
&&\cdot\, \exp\left(-\frac{\theta\sum_{x\in\Phi}\|x\|^{-\alpha}\fad{x}\tx{x}}{\|s\|^{-\alpha}}\right)\Bigg]\\\nonumber
&=&\sum_{k=0}^n {n\choose k} (-1)^k \suc{k+1}
\end{eqnarray}
with $\suc{k+1}$ given in~\eqref{eq:succ}.
\end{IEEEproof}

\begin{figure}[t]
\centering
\begin{tikzpicture}
\begin{axis}[xlabel={Sending probability $p$},ylabel={Probability $\outex{n}$},ymin=0,ymax=0.25,xmin=0,xmax=1,grid=both,
legend style={at={(0.98,0.98)}, anchor=north east, font=\footnotesize},
legend cell align=left,
yticklabel style={
        /pgf/number format/fixed,
        /pgf/number format/precision=2
}
]

\addplot plot[color=black,solid,no marks,style=thick]  table[x index=0,y index=1]  {poc.txt};
\addplot plot[color=black,solid,no marks,style=thick]  table[x index=0,y index=2]  {poc.txt};
\addplot plot[color=black,solid,no marks,style=thick]  table[x index=0,y index=3]  {poc.txt};
\addplot plot[color=black,solid,no marks,style=thick]  table[x index=0,y index=4]  {poc.txt};
\addplot plot[color=black,solid,no marks,style=thick]  table[x index=0,y index=5]  {poc.txt};

\node at(axis cs:0.42,0.22) {$n=1$};
\node at(axis cs:0.38,0.14) {$2$};
\node at(axis cs:0.38,0.105) {$3$};
\node at(axis cs:0.38,0.078) {$4$};
\node at(axis cs:0.38,0.059) {$5$};

\end{axis}
\end{tikzpicture}
\caption{Probability $\outex{n}$ that an outage is of length $n$ for different $p$. Parameters are $\theta=0.3$, $s=(1,0)$, $\lambda=1$, and $\alpha=3$.}\label{fig:poc}
\end{figure}
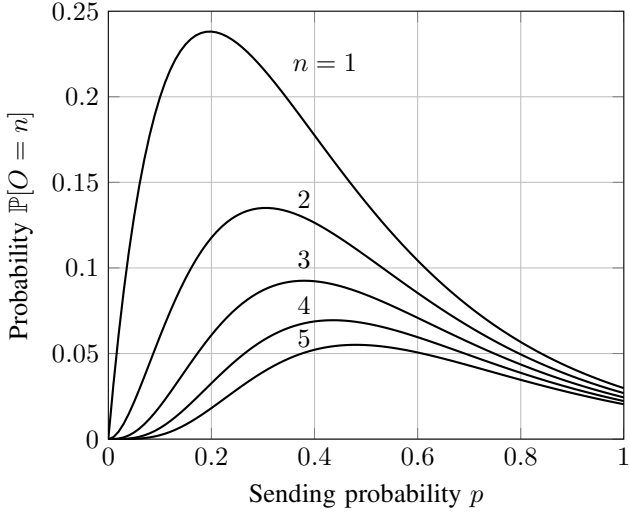

Fig.~\ref{fig:poc} shows the probability that an outage lasts $n=1,\dots,5$ slots. In this scenario, short outages are more likely than long ones, for all $p$ with the given outage probability. As expected, the peak of the probability is shifted  toward higher $p$ for increasing values of $n$, since higher interference correlation~$\rho$ implies longer outage~\cite{atiq19:eusipco}.

\section{Application to Erasure Correction Coding}
We now analyze the performance of erasure correction coding, focusing on random linear network coding~\cite{1705002}, for which $k$ source packets are encoded into $n$ coded packets. Here, each coded packet is formed by calculating a linear combination of the source packets with random coefficients. Successful decoding is only possible if the number of received packets $m$ is at least~$k$ and the corresponding coefficient vector matrix has rank $k$. Due to the random nature of the coefficient, the probability for the matrix to have rank $k$ increases with $m$ and with the size $q$ of the Galois field GF$(q)$ adopted for the coding. For successful decoding, the receiver obtains $k$ packets of information independent of~$m$.
The probability for successful decoding~is~\cite{5634159}
\begin{equation}
\decp =
\begin{cases}
0 & \mbox{if } m<k\\
\prod_{i=0}^{k-1}\left(1-\frac{1}{q^{m-i}}\right) & \mbox{else.}
\end{cases}
\end{equation}

Let us derive an expression for the throughput $\through$ of this code in our network setup. Let $\sucn{n}$ denote the number of successfully received packets in case $n$ packets are transmitted.
\begin{lemma}
The probability that out of $n$ transmitted packets a receiver is able to successfully detect any $k$ packets is
\begin{equation}
\suckn{n}{k}={{n}\choose{k}}\sum_{i=0}^{n-k} {{n-k}\choose{i}}\, (-1)^i\, \suc{k+i}\:.
\end{equation}
\end{lemma}
\begin{IEEEproof}
Similar to the proof of Lemma~\ref{lem:pmf-out}, the probability that any $k$ out of $n$ transmissions are successful and hence $n-k$ are in outage is
\begin{align}
\suckn{n}{k}&=\mathbb{E}\Bigg[\left(1-\exp\left(-\frac{\theta\sum_{x\in\Phi}\|x\|^{-\alpha}\fad{x}\tx{x}}{\|s\|^{-\alpha}}\right)\right)^k \\\nonumber
&\cdot\,\left(\exp\left(-\frac{\theta\sum_{x\in\Phi}\|x\|^{-\alpha}\fad{x}\tx{x}}{\|s\|^{-\alpha}}\right)\right)^{n-k}\Bigg]\,{n\choose k}\:.
\end{align}
Applying the binomial expansion yields the result.
\end{IEEEproof}

The throughput of random linear network coding is then
\begin{equation}
\through=\frac{k}{n}\,\sum_{m=k}^n\decp\:\suckn{n}{m}\:,
\end{equation}
where the fraction is the code rate and the sum gives the probability of successful decoding considering the channel.

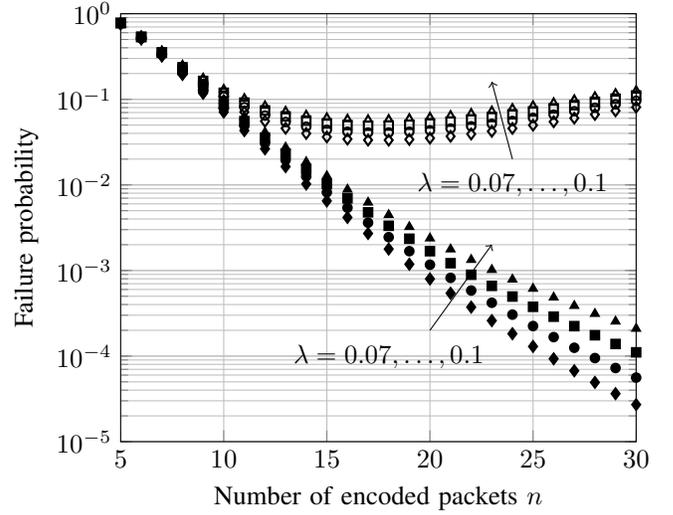
\begin{figure}[t]
\centering
\begin{tikzpicture}
\begin{axis}[xlabel={Number of encoded packets $n$},ylabel={Failure probability},ymin=10e-6,ymax=1,xmin=5,xmax=30,grid=both,
legend style={at={(0.98,0.98)}, anchor=north east, font=\footnotesize},ymode=log,
legend cell align=left
]

\addplot plot[color=black,mark=diamond,only marks,style=thick,mark options={scale=1,fill=black,solid}]  table[x index=0,y index=1]  {tradeoff_rlnc.txt};
\addplot plot[color=black,mark=diamond*,only marks,style=thick,mark options={scale=1,fill=black,solid}]  table[x index=0,y index=2]  {tradeoff_rlnc.txt};
\addplot plot[color=black,mark=o,only marks,style=thick,mark options={scale=0.8,fill=black,solid}]  table[x index=0,y index=3]  {tradeoff_rlnc.txt};
\addplot plot[color=black,mark=*,only marks,style=thick,mark options={scale=0.8,fill=black,solid}]  table[x index=0,y index=4]  {tradeoff_rlnc.txt};
\addplot plot[color=black,mark=square,only marks,style=thick,mark options={scale=0.8,fill=black,solid}]  table[x index=0,y index=5]  {tradeoff_rlnc.txt};
\addplot plot[color=black,mark=square*,only marks,style=thick,mark options={scale=0.8,fill=black,solid}]  table[x index=0,y index=6]  {tradeoff_rlnc.txt};
\addplot plot[color=black,mark=triangle,only marks,style=thick,mark options={scale=0.8,fill=black,solid}]  table[x index=0,y index=7]  {tradeoff_rlnc.txt};
\addplot plot[color=black,mark=triangle*,only marks,style=thick,mark options={scale=0.8,fill=black,solid}]  table[x index=0,y index=8]  {tradeoff_rlnc.txt};

\draw [<-] (axis cs:23,.16) -- (axis cs:24,.02);
\draw [<-] (axis cs:23,2e-3) -- (axis cs:20,2e-4);

\node at(axis cs:24,0.01) {$\lambda=0.07,\dots,0.1$}; 
\node at(axis cs:18,1e-4) {$\lambda=0.07,\dots,0.1$}; 

\end{axis}
\end{tikzpicture}
\caption{The failure probability of a typical link in the network for different coding rates $5/n$. Filled marks are for uncorrelated interference, while open marks are for correlated interference. The sending probability is $p=n/30$; parameters are $\theta=1$, $s=(1,0)$, $\alpha=4$, and $q=2$.}\label{fig:tradeoff}
\end{figure}

Fig.~\ref{fig:tradeoff} shows the failure probability of a typical link, i.e., the probability that a node is unable to decode a set of packets. We assume that $k=5$ data packets are encoded by random linear network coding into $n$ coded packets, out of which a subset of $m$ packets reaches the receiver. This implies that the overall sending probability and in turn interference increases linearly with $n$, as we assume that all links in the network adopt the same code. The plot shows traces for both correlated and uncorrelated interference. As can be seen, correlation degrades the performance of the code significantly and introduces an optimal amount of redundancy in terms of decoding probability (around $n=17$), while for uncorrelated interference a higher $n$ is beneficial within the considered interval. This optimal $n$ arises from the following tradeoff: For low values of $n$, redundancy is too small for the channel conditions and hence many packets cannot be recovered. Thus, when increasing $n$ the throughput increases up to its maximum. However, for increasing $n$, also the interference gets higher reducing the reception probability; at some point this outweighs the additional benefit of the code and in turn the decoding probability decreases. An important reason is that interference correlation diminishes the gain of higher $n$.

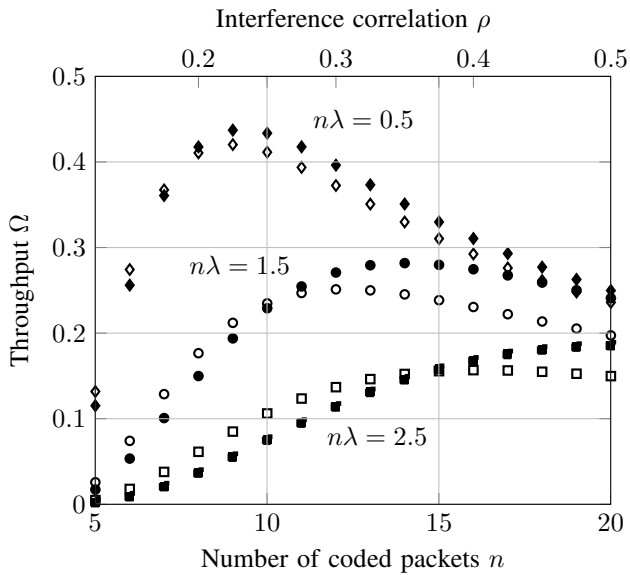
\begin{figure}[t]
\centering
\begin{tikzpicture}
\begin{axis}[xlabel={Interference correlation $\rho$},ylabel={Throughput $\through$},ymin=0,ymax=0.5,xmin=0.125,xmax=0.5,grid=none,axis x line*=top,axis y line=none,
legend style={at={(0.98,0.98)}, anchor=north east, font=\footnotesize},
legend cell align=left
]

\addplot plot[color=black,mark=diamond,only marks,style=thick,mark options={scale=1,fill=black}]  table[x index=1,y index=2]  {through_norm_rlnc.txt};
\addplot plot[color=black,mark=diamond*,only marks,style=thick,mark options={scale=1,fill=black}]  table[x index=1,y index=3]  {through_norm_rlnc.txt};
\addplot plot[color=black,mark=o,only marks,style=thick,mark options={scale=0.8,fill=black}]  table[x index=1,y index=4]  {through_norm_rlnc.txt};
\addplot plot[color=black,mark=*,only marks,style=thick,mark options={scale=0.8,fill=black}]  table[x index=1,y index=5]  {through_norm_rlnc.txt};
\addplot plot[color=black,mark=square,only marks,style=thick,mark options={scale=0.8,fill=black}]  table[x index=1,y index=6]  {through_norm_rlnc.txt};
\addplot plot[color=black,mark=square*,only marks,style=thick,mark options={scale=0.8,fill=black}]  table[x index=1,y index=7]  {through_norm_rlnc.txt};

\node at(axis cs:0.32,0.45) {$n\lambda=0.5$}; 
\node at(axis cs:0.23,0.28) {$n\lambda=1.5$}; 
\node at(axis cs:0.33,0.08) {$n\lambda=2.5$}; 

\end{axis}
\begin{axis}[xlabel={Number of coded packets $n$},ylabel={Throughput $\through$},ymin=0,ymax=0.5,xmin=5,xmax=20,grid=both,xtick={5,10,15,20}]
\end{axis}
\end{tikzpicture}
\caption{The throughput of a typical link in the network for varying the coding rate $5/n$. The sending probability is $p=n/20$ and hence interference correlation is $\rho=n/40$, which is shown as second x-axis on top. Filled marks are for uncorrelated interference (top x-axis does not apply), while open marks are for correlated interference. The intensity $\lambda$ is chosen to keep $\lambda\txp$ and in turn the outage probability constant. Parameters are $k=5$, $\theta=1$, $s=(1,0)$, $\alpha=3$ and $q=2$.\vspace{-1.2mm}}\label{fig:through}
\end{figure}

In order to analyze the impact of interference correlation in more detail, we plot the throughput $\through$ of a node over the number of coded packets $n$ in Fig.~\ref{fig:through}. A second $x$-axis on the top shows the interference correlation $\rho=n/40$, as we assume $p=n/20$. This axis only applies to the open marks that depict $\through$ for correlated interference.
The success probability $\suc{1}$ is kept constant by keeping $p\lambda=n\lambda/20$ constant.
The plot shows that in the low interference regime ($n\lambda=0.5$) the optimum is pronounced stronger as compared to Fig.~\ref{fig:tradeoff} and even exists in uncorrelated interference traces. The reason is that throughput decreases stronger with $n$ than the decoding probability, as the coding rate $k/n$ is also impacting this decrease. Intuitively speaking, if already more than $k$ packets have been received, the decoding probability will not increase much by receiving further packets, but these are consuming bandwidth and in turn reduce the throughput. It is important to note that this is not due to an increase of interference, as $\lambda\txp$ is kept constant.
When evaluating the impact of interference correlation on throughput, we can see that for lower $n$ it depends on the scenario whether having correlation yields higher or lower throughput, similar to Fig.~\ref{fig:tradeoff}. For high $n$, however, correlation is decreasing throughput in all cases, as for this high number of encoded packets it is advantageous to have many independent chances of receiving packets rather than the ``all-or-nothing'' situation of high correlation.

Finally, note that the decrease of throughput for high $n$ as shown in Fig.~\ref{fig:through} is only moderate due to a constant success probability $\suc{1}$. In a similar plot with constant $\lambda$, i.e., with varying interference, the decay of throughput with increasing $n$ would be stronger.

\section{Conclusions}

We derived and analyzed the outage and success duration in Poisson networks by applying tools from stochastic geometry.
Based on these results, we analyzed the performance of random linear network coding in terms of decoding probability and throughput. 
Results show that an optimal number of coded packets exists at which enough coded packets are transmitted for good performance while limiting the interference to a moderate level.
Furthermore, we show how interference correlation can increase or decrease the decoding probability depending on the network scenario. Potential applications can be found, e.g., in the field of industrial sensor networks. Further steps involve the practical exploitation of our results to optimize the coding rate in a sensor network within an industrial facility.

The general insights, although investigated for particular modeling assumptions here, qualitatively { generalize} to a broader range of networks that exhibit interference correlation (e.g., Mat\'ern networks~\cite{schilcher19:matern}).

\end{document}